# Uncovering hidden protein conformations with high bandwidth nanopore measurements

Kyril Kavetsky[1,2], Sabine Hong[1], Chih-Yuan Lin[1], Roger Yang[1], Marija Drndić[1]*

[1]*Department of Physics and Astronomy, University of Pennsylvania, Philadelphia, PA 19104, USA*

[2]*Department of Material Science and Engineering, University of Pennsylvania, Philadelphia, PA 19104, USA*

\*: corresponding author
Email: drndic@physics.upenn.edu

**ABSTRACT**

Advanced nanopore measurements allow structural probing of molecules with high spatial and temporal resolution. We report high signal-to-noise, 1-10 MHz bandwidth, translocation measurements of the multi-state folding of heme protein cytochrome c in KCl solution through optimally-designed silicon nitride pores of 2.3 – 3.3 nm diameter and 3.6 - 3.8 nm effective thickness, and an optimal concentration of a denaturant (Gdm-Cl). The pore diameter is slightly smaller than the protein's size, forcing the protein to squeeze through the pore. The sufficiently large pore thickness allows enough time for protein probing at an applied field of ~ 250 kV/cm. Through Bayesian Information Criterion score analysis, current blockades reveal six distinct levels, attributed to specific protein states. We calculate the transition probabilities between the states and the conditional probabilities of the protein leaving the pore from each state. We validate the model by simulating events and comparing them to experimental data.

**MAIN TEXT**

Resolving the primary structures of biomolecules, such as nucleobase sequences in DNA and amino acids in proteins, has been a longstanding challenge for solid-state nanopores.[1,2] Protein-based nanopores, functionalized with motor enzymes for high spatial control, are advantageous for these purposes.[3] Solid-state pores have been highly successful in characterizing higher-order structures in DNA,[4–6] RNA,[7–9] and proteins,[10–14] owing to the customizability of devices and experimental conditions. Unlike protein nanopores, solid-state pores are particularly interesting because they can be tuned for sensitivity to both native-state and denatured forms of analytes, and secondary structures can thus be inferred from translocation measurements.[15,16] For example, solid-state nanopores have been used to detect the calcium-induced conformation change in calmodulin, as evidenced by marked differences in both current blockade amplitude and dwell time.[17] Yusko *et al.* investigated the rotational dynamic of proteins during their translocations through nanopores.[13] By modifying the nanopore surface, they achieved the characterization of individual proteins and protein complexes in terms of their size, shape, and dipole moments. Recently, our group demonstrated that the low-noise nanopore platform can distinguish transfer RNA (tRNA) molecules differing by only a single-nucleotide substitution, which advances its use for probing the conformational dynamics of RNA molecules associated with human diseases.[8]

There is an optimal, relatively narrow range of nanopore diameters and applied electric fields, for which protein conformations can be detected. High bandwidth (> 1 MHz) nanopore measurements also allow high temporal resolution,[12,18,19] which is crucial because protein translocation is fast and detection is challenging.[20] This can enable observations of dynamic reconfiguration of the molecules, as in the case of multi-state folding proteins. This is illustrated by cytochrome c (cyt c), a small (~3.4 nm across) mitochondrial protein which plays roles in cellular respiration and apoptosis.[21] To signal apoptosis, the protein must pass through porins in the mitochondrial membrane, whose openings have been reported to range in diameter from 2 to 3 nm. Translocation measurements of this protein found that when nanopore diameter was decreased below 3.0 nm, higher voltages were required to observe translocations with prolonged dwell times compared to the native protein translocation in larger pores, as the protein had to squeeze through the narrow pore.[22] For diameters < 2 nm, the protein translocated unfolded while for a larger 4-nm-diameter pore,[22] two-level signals were reported using higher bandwidth electronics.[23]

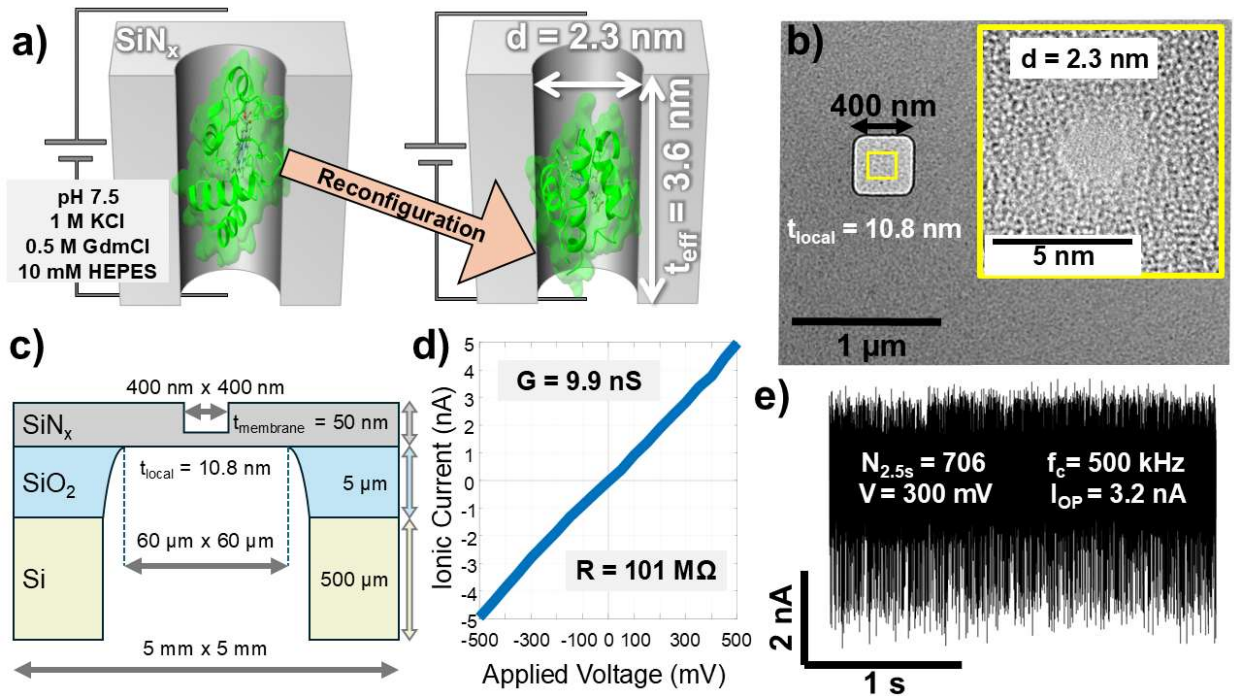

**Figure 1**. **Experimental setup**. (a) Schematic of cyt c translocation through a $SiN_x$ nanopore device. Cyt c reconfigurations are schematically portrayed as distortions in the protein's shape. Three-dimensional renderings of cyt c were generated using ChimeraX. (b) TEM image of a silicon nitride membrane locally thinned to ~ 11 nm. Inset is a TEM image of a nanopore with a diameter d = 2.3 ± 0.2 nm. (c) Schematic of the device substrate prior to TEM drilling. The pore is formed in the thin membrane etched from a freestanding 50 nm layer of LPCVD $SiN_x$ via EBL and RIE. A 60 μm square region of silicon nitride is supported by a 5 μm $SiO_2$ on 500 μm of Si, which were removed by buffered oxide and KOH wet etches, respectively, to isolate the freestanding $SiN_x$ membrane. (d) Current vs. voltage (I-V) curve of the 2.3 nm diameter pore. The conductance is determined by calculating the slope of the I-V curve, and the calculated conductance is used to estimate the pore thickness. (e) 2.5 second long current trace of cyt c translocation at applied voltage V = 300 mV. The open-pore current ($I_{OP}$) at this voltage is 3.2 nA. The current trace was filtered using a low-pass Bessel filter to a cutoff frequency of 500 kHz. An automated threshold search detected 706 events in this trace.

In this study, we use a high bandwidth (up to 10 MHz) solid-state nanopore platform to observe the high-speed folding dynamics of cyt c during nanopore translocation. High signal-to-noise ratio (SNR) measurements (SNR > 6 for d = 2.3 nm pore with V = 300 mV when a 200 kHz cutoff filter is used), optimized nanopore size, and robust signal processing reveal previously overlooked protein-nanopore interactions at biologically relevant length scales. We demonstrate this by designing, fabricating, and implementing a constricting nanopore platform and measuring the resistive pulses of the protein-pore interactions, then identifying distinct current levels and performing basic characterizations based on the data generated. We calculate a Bayesian Information Criterion (BIC) score and observe six distinct states in a 2.3 nm diameter pore at 300 mV and 500 mV applied voltages when BIC score is minimized. We characterize each event as a series of configurations to uncover patterns in the protein's folding behavior. This includes transition probabilities among the states observed at 300 mV based on a large total number of translocation events (>11,000) and total number of states within those events (>21,000). This nanopore-based method, when combined with additional structural characterization techniques, can provide detailed insight into the structures of a wide variety of confined proteins and their associated reconfiguration dynamics.

**Figure 1a** is a schematic of protein[24] translocation through a silicon nitride ($SiN_x$) nanopore with device details. We use nanopore diameters (2.3 – 3.3 nm) smaller than that of cyt c (~3.4 nm), such that the pore acts as a constriction. First we observe that, as found previously,[22] a nanopore of diameter ranging from 2 nm to 3 nm promotes protein squeezing and folding, as the applied voltage provides the driving force for protein transport. Nanopores larger or smaller than this range were found to either allow the protein to translocate in its native state, or fully unfolded, respectively. The narrow constriction (diameter, d = 2.3 ± 0.2 nm nm for Device 1, **Figure 1**) and pore thickness exceeding the protein diameter ($t_{eff}$ = 3.6 nm, **Figure 1**) allow translocation times long enough for the protein to be efficiently probed with our high-bandwidth setup, while the current signal is still large enough given these pore dimensions. We observe relatively long characteristic dwell times (a range from ~ 20 to 170 µs for voltage range 100 to 500 mV for the 2.3 – 3.3 nm pore diameter, **Tables S1, S2** and **Figures S4, S6**), which maximize the number of reconfigurations cyt c undergoes during translocation, generating more data regarding its folding behaviors.

**Figure 1b** is a transmission electron microscope (TEM) image of a 2.3-nm-diameter nanopore (Device 1) as drilled in the locally thinned $SiN_x$ membrane. Silicon chips having 5-µm thick oxide (as illustrated in **Figure 1c**) are utilized to minimize the noise when operating with high-bandwidth electronics and maximize the SNR. The effective thickness, $t_{eff}$, is calculated from eq S1 based on the measured open pore conductance (**Figure 1d**), electrolyte conductivity, and TEM-measured diameter. The driving voltage must be carefully selected, as it also must balance a manageable translocation speed, driving force, and SNR. We tested driving voltages up to 500 mV and found that ~ 300 mV results in pronounced folding behavior. A second 3.3-nm-diameter pore (Device 2) was drilled in a $SiN_x$ membrane with similar effective thickness, $t_{eff}$ = 3.8 nm (**Figure S1**). In the rest of this letter, we discuss results from Devices 1 and 2.

Given the possible signal magnitudes from cyt c, the highest cut off frequency at which we were able to identify events was ~ 1 MHz for both devices, limited by the total noise from both the chip and the amplifier.[18] Additional ion current *vs.* time traces recorded at different voltages for two nanopores at bandwidths 1 and 10 MHz, respectively, are shown in **Figures S2** and **S9.** We used two different amplifier setups (**Table S1**) from Chimera (Device 1) and Elements (Device 2). The measured root-mean squared (RMS) current noises were $I_{rms}$ = 1.3 $nA_{rms}$ for Device 1 and 0.8 $nA_{rms}$ for Device 2 at 1 MHz, respectively, and 2.58 $nA_{rms}$ for Device 2 at 10 MHz. The $I_{rms}$ at lower cutoff frequencies are listed in **Figure S9** for Device 2. **Table S4** compares the current noise ($I_{rms}$) measured with a 10 MHz amplifier used in this work for Device 2 to values reported in previous studies (that range from 1.1 to 3.1 nA),[19,25] obtained from unfiltered (10 MHz bandwidth) traces. The $I_{rms}$ value for Device 2 in this work (2.58 $nA_{rms}$) is similar to the value reported in the first 10 MHz recoding using $SiN_x$ nanopores on glass chips, 2.5 $nA_{rms}$.[25]

The translocation trace in **Figure 1e** reproduces well the values of the open pore conductance ($\approx$ 10 nS) and mean fractional blockades (~ 47% at 300 mV) reported by translocating this protein through an almost identical pore (2.5 nm diameter and 3.4 nm effective thickness, compared to 2.3 nm diameter and 3.6 nm effective thickness) in a previous measurement.[22] The higher bandwidth used here (1 MHz) allows the detection of much shorter features in the translocation signals. It is also interesting to compare our nanopore properties with the mitochondrial membrane, where the potential difference that drives cyt c expulsion through the voltage-dependent anion channel (VDAC) is about 50 mV, which corresponds to a local electric field of 100 kV/cm.[26] In our case, we use 300 mV across a ~11 nm membrane to observe translocation events with multiple levels, corresponding to an electric field ~ 250 kV/cm.

**Figures 2a** and **2b each** show a set of scatter plots of mean fractional blockades as a function of dwell times at 300 mV and 500 mV, respectively, for Device 1 and Device 2 (TEM images are shown in **Figure S1**). The cutoff frequency used for these plots is 500 kHz. More current traces for both devices are shown in **Figures S2** and **S9** and the individual scatter plots with indicated numbers of events are displayed in **Figure S4**. The larger diameter nanopore device (Device 2) shows smaller mean fractional blockades and the dwell time distribution spreads a bit more towards larger times. **Table S2** lists the open pore currents, absolute and relative mean current blockades, and characteristic dwell times for both devices at a range of voltages.

**Figure 2d** shows examples of cyt c translocation signals with step averages shown between changepoints. Device 1 (2.3 nm pore) signals were filtered to a cutoff frequency of 500 kHz, and Device 2 (3.3 nm pore) signals were filtered to 200 kHz. All translocation measurements for both devices were performed in 1 M KCl solution at pH 7.5 with added denaturant (0.5 M Gdm-Cl) and 10 mM HEPES, as previously optimized using an almost identical nanopore to Device 1.[22] Automated event detection was conducted using a threshold search, and step changes in translocation event signals were detected using a differential residual squared error optimization method (**Figure S3**), which we have previously used to characterize folding in double-stranded DNA at 10 MHz bandwidth.[19] An overview of the event and step-change detection processes is shown in the **Supporting Information** (**Section S2**). **Figure 2c** shows histograms comparing ionic current levels averaged between changepoints with those averaged over the entire translocation events. The presence of multiple populations among step averages implies distinct folding states, which is not apparent if only mean current blockade is considered. **Figure 2e** summarizes the dataset acquired from a 2.3 nm diameter pore (Device 1) at V = 300 mV. The number of events in the dataset is 11507 with 47% of events having two steps. Events with more than one step are prevalent, 63% and 52% of events have two or more steps at 300 mV and 500 mV, respectively.

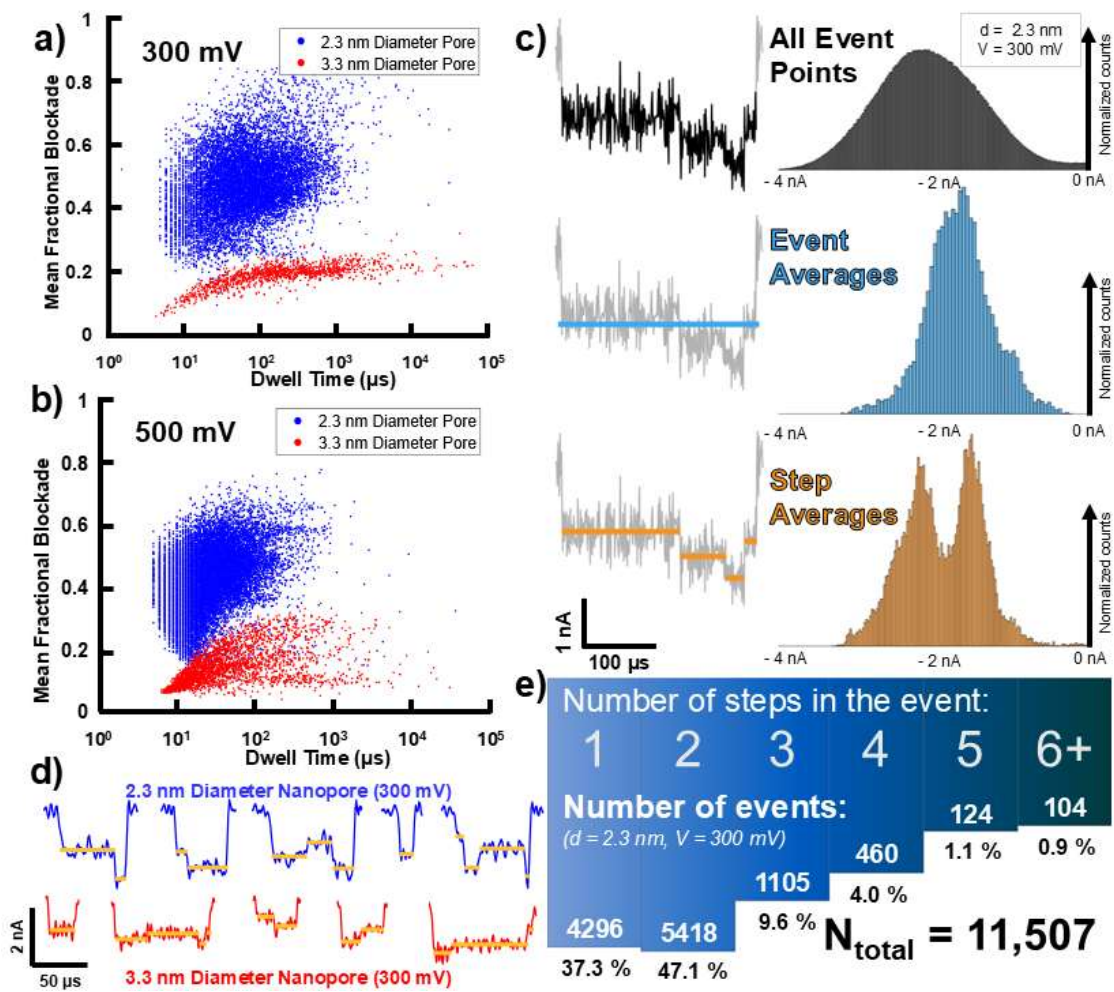

**Figure 2. Cytochrome c translocation metrics.** (a, b) Scatter plots of fractional current blockade vs. dwell time at 300 mV applied voltage for a 2.3 nm and 3.3 nm diameter nanopore, respectively, and effective thickness $t_{eff}$ ~ 4 nm. (c) A breakdown of ionic current depth distribution across multiple levels of event analysis. The all-points histogram represents the distribution of all sampled currents during translocation events measured using Device 1 at 300 mV, as indicated with the example event trace (left). The event-averaged distribution represents the distribution of the ionic current blockade averaged per event, as indicated by the blue line extending over the mean of the example event. The step-averaged distribution considers the average ionic current blockade of the event signal between each set of changepoints, as represented by the orange lines showing the average of each step in the example signal. (d) Selected cyt c translocation event signals for Device 1 (500 kHz filtered) and Device 2 (200 kHz filtered) at 300 mV. Events vary in blockade depth, dwell time, and in the number of folding transitions occurring during the event. Step averages between changepoints in the event signal are shown in orange. (e) Percentage and number of events *vs.* number of steps in the event (1 to 6+) from Device 1 at 300 mV and cutoff frequency 500 kHz. The total number of events, $N_{total}$, is 11,507. One-step events (no conformation changes detected) comprise 37.3% of the total number of events, while most of the events (7,211) have at least one folding transition. The bar graph shown is in an inverted log scale.

A discussion on the relationship between applied electric field and number of steps in the events during cyt c translocation is provided in **Section S5**, where we determine that relatively high (500 mV) and low voltages (100 mV) reduce the relative prominence of multi-state folding events. The characteristic dwell time decreases from 69 to 18 microseconds as voltage is increased from 300 mV to 500 mV (**Figure S6**). Some short-lived steps are also more easily missed at higher voltage due to bandwidth limitation.

We performed measurements with sampling rates of 40 MHz with Device 2. A summary of the measurement conditions for all high-bandwidth measurements presented in this work is provided in **Table S1**. The mean current blockades vs dwell times for events at 300 mV and 500 mV are shown in **Figure S4**. A summary of the translocation metrics for two devices for all measured voltages is given in **Table S2.**

Each event was classified based on the mean current blockade of each step detected within the event. Events with multiple steps detected (two or more) were used to form a Hidden Markov Model (HMM) for determining patterns in the folding behavior.[27,28] Preparation of the data for forming the basis of the HMM is described in detail in **Section S2**. The total number of states available was determined by a Bayesian Information Criterion (BIC) optimization, although it should be noted that this result is also dependent on the cutoff frequency used to filter translocation signals. Additional information about the effect of cutoff frequency on BIC optimization is provided in **Section S2.3**. Importantly, six states are consistently detected for a wide range of cutoff frequencies from 100 kHz to 500 kHz (see **Figure S5**), showing that the number of states is not an artifact of the choice of cutoff frequency. The results for 500 mV are shown in **Figure S5b** and lead to similar conclusions presented here.

**Figure 3a** shows the results of a multi-run assessment of BIC scores for models with a set number of available states at 300 mV for Device 1. For a cutoff frequency of 200 kHz, BIC scores are minimized when a 6-state model is used, indicating that this model produces the best fit at this frequency. The BIC score for 6 states is also minimized at 500 mV for this cutoff frequency and pore size (**Figure S5d**). We find that 200 kHz provides a favorable balance of noise performance (SNR > 6 at 200 kHz) while still resolving shorter-lived states. Here, SNR is defined as the ratio between the mean current blockade divided by the RMS noise in the open pore current. Each state corresponds to a distribution of current blockade values, which the algorithm uses to assign a state to each ionic current step in multi-step events. **Figure 3b** shows the distributions of these current blockades for each detected state. A Viterbi algorithm[29] is then used to assign the most likely sequence of configuration states to all events based on the mean ionic current in each step. As shown in **Figure 5c,** the mean current blockade that has been assigned to an individual configuration state varies from 1.1 nA to 2.3 nA (corresponding to fractional blockades of 0.35 and 0.73, respectively). These observed blockades are much larger than the estimated blockade caused by a fully unfolded protein (current change of 0.5 nA and fraction of 0.16) with a cylinder model,[30,31] assuming diameter of 1 nm for a peptide chain. This also suggests that cyt c protein does not fully denature and still holds its folding structure while translocating through the nanopore under these experimental conditions (0.5 M GdmCl).[22] **Figure 3c** shows ionic current traces of example multi-step events with the means of each step indicated and labeled according to the configuration state assigned by the Viterbi algorithm. Additional details about how the Viterbi algorithm is applied are available in **Section S4.4**.

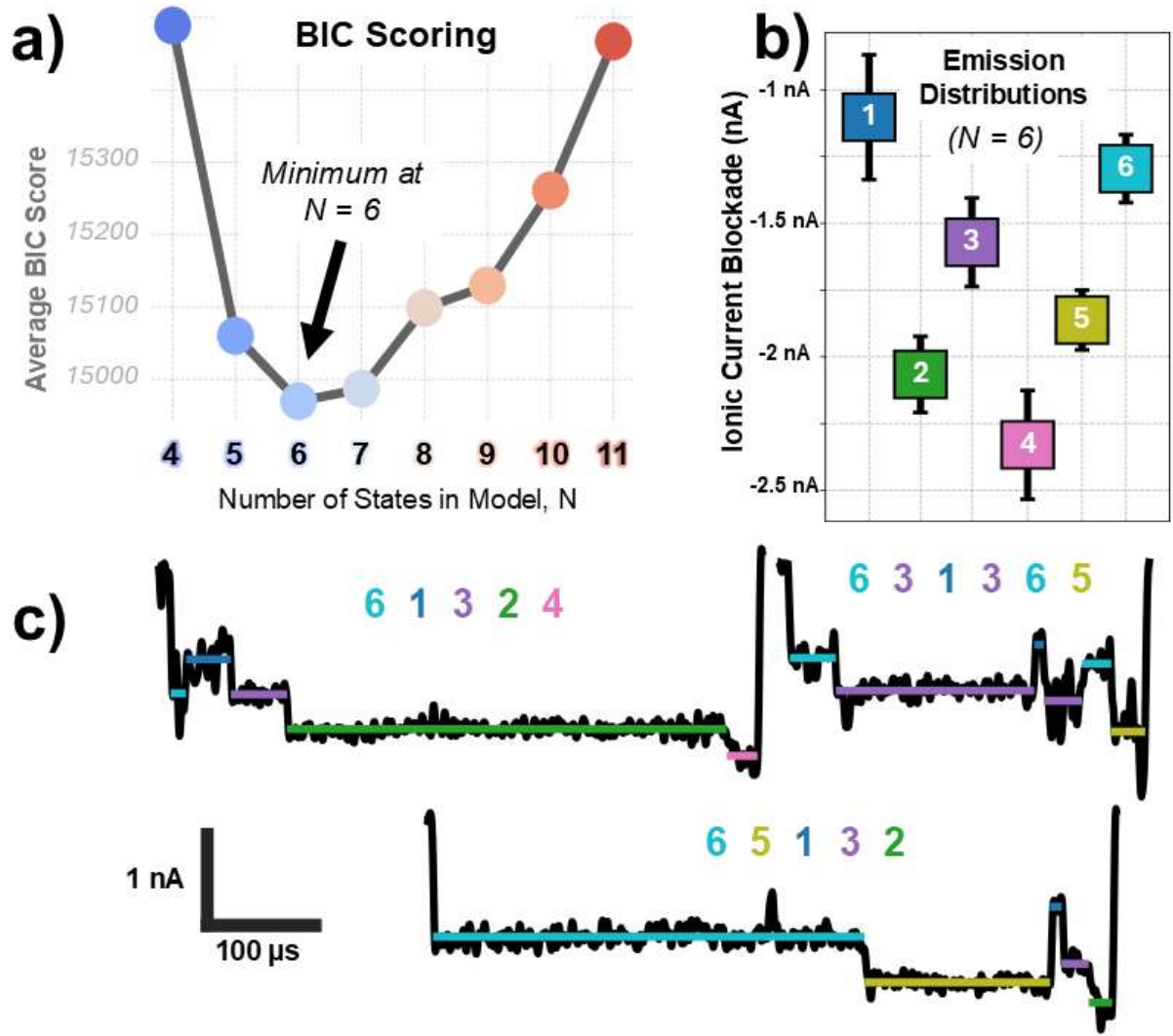

**Figure 3**. **Model Selection and Assignment of States.** Mean ionic current blockade values for each step in the dataset were processed by a Viterbi algorithm. **a)** The average BIC score from each model after 500 iterations. Average BIC score is minimized for models of 6 states. **b)** Current blockade profiles for each state in a 6-state Markov model are shown as a box plot, with each node representing the mean and each stem extending one standard deviation. **c)** Example multi-step events with associated state assignments. The mean of each step in the event is coded with the color corresponding to the state assigned to it.

**Figure 4** presents the results of the HMM for the 2.3 nm diameter pore (Device 1). The frequency of assignment for each state is detailed in **Figure 4a**. For 100% of events, the first step is assigned with State 6, indicating that this state corresponds to the non-constricted configuration of the molecule upon pore incidence. We calculate that the access resistance contributes approximately 17% of the total pore resistance (**eq S1**). This is smaller than the observed fractional current blockade of 35% in State 6, suggesting that State 6 corresponds to protein residence. To confirm this, we also performed numerical modeling of the resistive pulse expected at the pore, which is shown in detail in **Section S6** of the **Supporting Information (Figure S7)**. The result suggests that achieving such a high blockade requires the protein to penetrate slightly into the nanopore and/or carry more charges. The high charges carried by the protein may also contribute to the higher blockade (**Figure S8**), as reported previously.[32]

The results of the HMM are graphically summarized in **Figure 4c**. A matrix of all calculated transition probabilities, rounded to the nearest percent, is shown in **Table S3**. Main diagonal elements correspond to "self-transitions", where the model does not assign a new state following a detected changepoint. These self-transitions are uncommon and could be the result of noise and limitations of the changepoint detection algorithm. To further validate our model, we conversely use these calculated transition probabilities to generate simulated translocation events and compare them with measured events. **Figure 4d** shows simulated ionic current traces and compares them with real translocations from the dataset shown in **Figure 2**. This is achieved using the respective probabilities of state transitions (**Figure 4c**) and durations (**Figure 5b**) to create a series of steps, tuning the noise level to approximately match the input data, and locating an event in the dataset with similar behavior. A detailed explanation of how these events are simulated is described in **Section S4** in the **Supporting Information**.

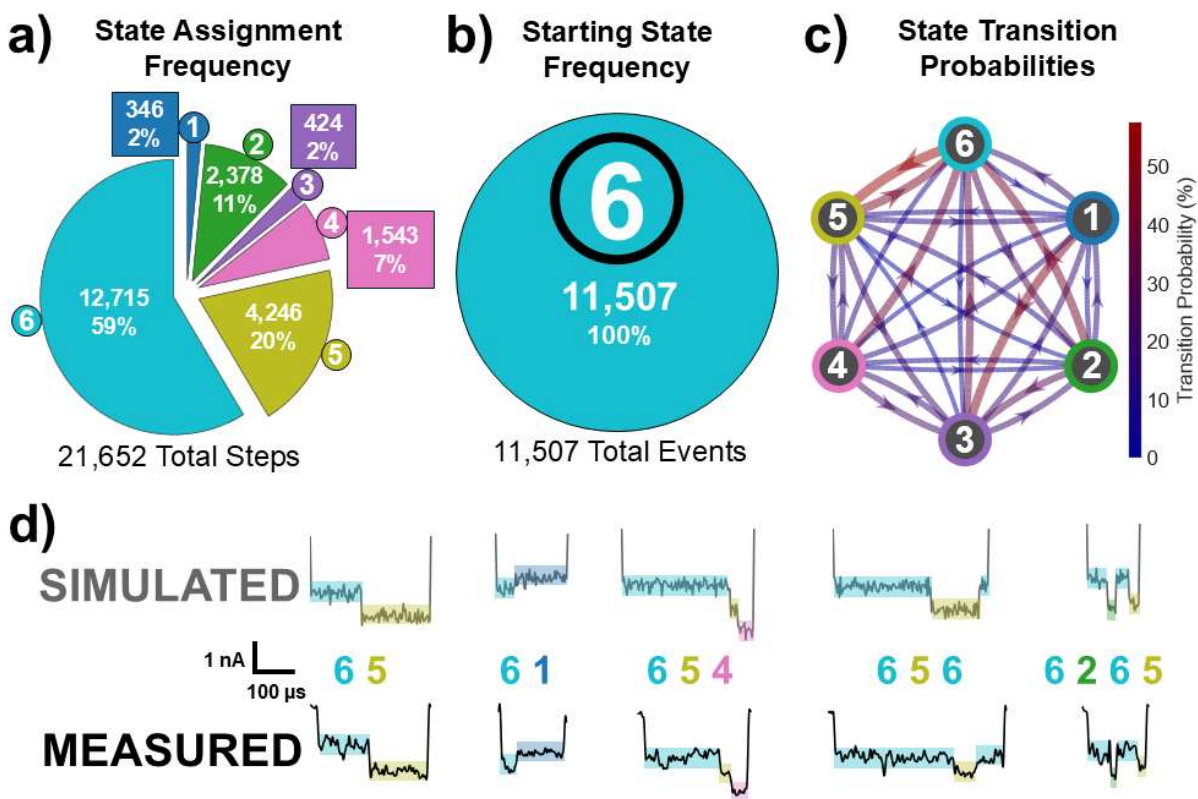

**Figure 4. Hidden Markov Model Statistics.** Summary of statistics of protein configuration behavior as observed from the HMM. (a) The frequency of state assignment among all events for each of the six states, presented with the associated percentage. (b) Pie graph of starting state probabilities. The Viterbi analysis shows that all events begin with the protein in the same configuration state (State 6). This indicates that State 6 corresponds to the unconstricted configuration of the suspended protein that initially reaches the pore. (c) Markov Chain visualization of the reconfiguration patterns observed. Each transition has an associated probability. (d) Simulated cyt c translocation events based on results of the HMM. The simulated events (above) are compared with real events from the data set (below).

From the Viterbi state assignments across multi-step events in the dataset, we characterize the states according to their respective termination probabilities (conditional probability that the event will end if the protein is in a particular state), durations, and mean current blockade. This analysis allows us to compare the relative behaviors of these protein configurations. The event termination probability (**Figure 5a**) indicates the protein's ability to leave the nanopore in this state. For example, an event has a 95% chance of ending if the protein is in State 2 compared to 36% if the protein is in State 6. Furthermore, analysis of state duration (**Figure 5b**) shows the longevity of each state. For example, we find State 4 has the longest duration (~70 μs characteristic dwell time, τ), and State 1 (τ ~ 26 μs) to have the shortest. More details about the methods we employ for determining characteristic durations for each state are provided in **Section S3.2** in the **Supporting Information**. **Figure 5c** compares the average ionic current blockade, with the corresponding blockage fractions of open-pore current, for each state.

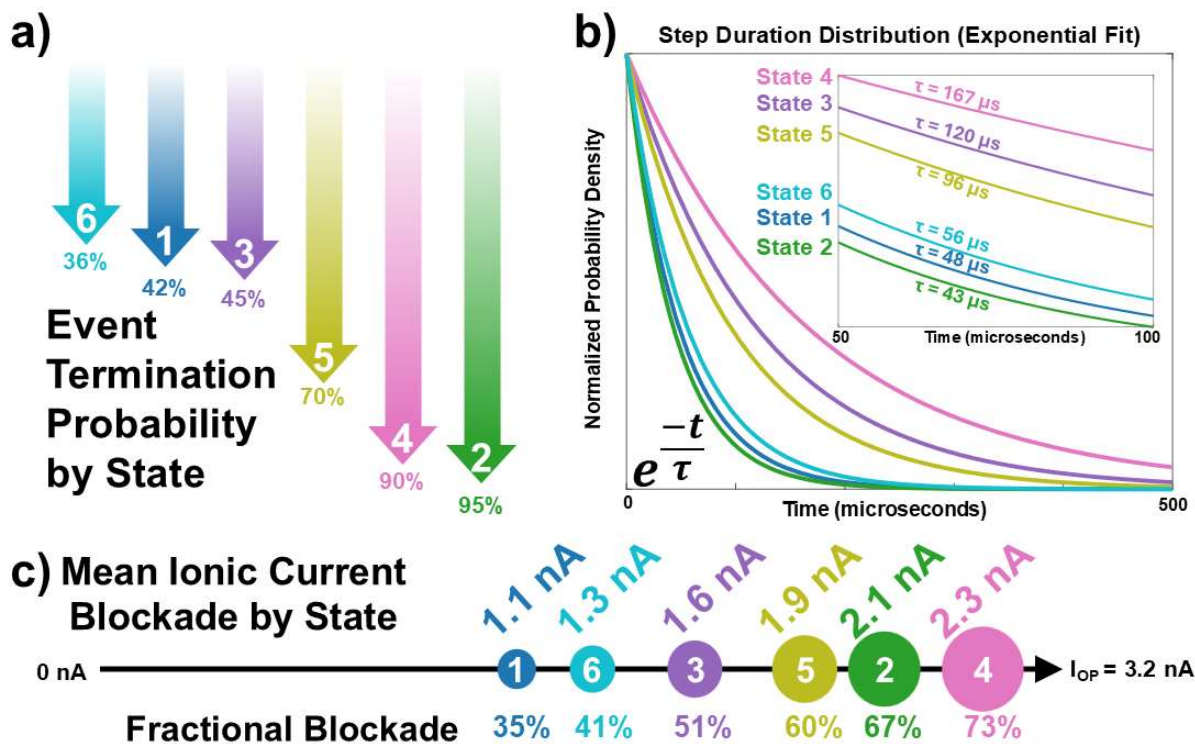

**Figure 5**. **HMM Analytics Results.** Results of the HMM allow for certain inferences to be made regarding the configuration states of the protein. (a) Across multi-step events, we calculate the conditional probability for each state of event termination (instead of a transition to another state). This metric indicates the ease with which the protein exists the pore in each state. (b) Exponential decay fits of state duration distributions allow us to characterize the longevity of each state during the translocation. The characteristic duration τ reflects the overall longevity of the state. (c) The mean ionic current blockade for each state across all assignments reflects the volume excluded by the protein when in each configuration.

Multiple characterizations based on the current blockade levels and durations of the states can be considered together to further speculate into the protein's overall shape and behavior in each state. For example, State 1 has the lowest current blockade, and the second-lowest characteristic duration and event termination probability. It is therefore possible that the protein assumes a highly denatured unfolded configuration in State 1 and reconfigures quickly into a more compact structure such as State 3. On the other hand, State 2 has the shortest characteristic duration, the second-highest mean current blockade, and the highest event termination probability. This indicates that State 2 could correspond to a constricted conformation which quickly passes through the pore or otherwise relaxes into a less-squeezed conformation. In contrast to States 1 and 2, State 5 is a relatively frequently assigned configuration which ranks moderately in each of these categories.

In summary, using high bandwidth (1-10 MHz) measurements of optimally chosen nanopore dimensions and voltages, we discover and characterize six states within the translocation signals of cyt c. In addition, we describe how these levels are identified, and we construct a model of protein folding behavior using only ionic current measurements. The model gives no direct information on the folding structure without *a priori* information, but it allows for inferences to be made about conformations and behaviors of the protein present in each state. A model for cytochrome c reconfiguration upon translocation through a nanopore constriction could be used to improve simulations of molecular biological systems during cellular respiration. Furthermore, this technique can be extended to study the folding behaviors of other molecules through constrictions. In such cases, folding pattern information can be combined with structural characterization to form more comprehensive models of molecular dynamics in biological systems.

ASSOCIATED CONTENT

The **Supporting Information** is available free of charge via the Internet at http://pubs.acs.org. Nanopore fabrication (Section 1), data analysis algorithm (Section 2), characterization of configuration states (Section 3), simulation using HMM results (Section 4), translocation metric across voltages (Section 5), Numerical model (Section 6), noise analysis (Section 7), and current traces from 10MHz amplifier (Section 8).

AUTHOR INFORMATION

**Corresponding Author**

*Email: drndic@physics.upenn.edu

ACKNOWLEDGMENT

This work was partially funded by NIH R21 HG011120 and NIH R01 HG01241 grants. The Si-based chips were fabricated in the Singh Center for Nanotechnology at the University of Pennsylvania, supported by the NSF grant no. NNCI-1542153. We acknowledge the use of TEM instrumentation supported by the NSF through the University of Pennsylvania Materials Research Science and Engineering Center (MRSEC) DMR-1720530. Molecular graphics and analyses were performed with UCSF ChimeraX, developed by the Resource for Biocomputing, Visualization, and Informatics at the University of California, San Francisco, with support from National Institutes of Health R01-GM129325 and the Office of Cyber Infrastructure and Computational Biology, National Institute of Allergy and Infectious Diseases. The authors acknowledge Dr. Srilahari Namani, Dr. Federico Thei, Dr. Filippo Cona, and Serge Kaddoura for their assistance with measurements, analysis, and helpful discussions throughout various stages of the completion of this work.

# REFERENCE

(1) Alfaro, J. A.; Bohländer, P.; Dai, M.; Filius, M.; Howard, C. J.; Van Kooten, X. F.; Ohayon, S.; Pomorski, A.; Schmid, S.; Aksimentiev, A.; Anslyn, E. V.; Bedran, G.; Cao, C.; Chinappi, M.; Coyaud, E.; Dekker, C.; Dittmar, G.; Drachman, N.; Eelkema, R.; Goodlett, D.; Hentz, S.; Kalathiya, U.; Kelleher, N. L.; Kelly, R. T.; Kelman, Z.; Kim, S. H.; Kuster, B.; Rodriguez-Larrea, D.; Lindsay, S.; Maglia, G.; Marcotte, E. M.; Marino, J. P.; Masselon, C.; Mayer, M.; Samaras, P.; Sarthak, K.; Sepiashvili, L.; Stein, D.; Wanunu, M.; Wilhelm, M.; Yin, P.; Meller, A.; Joo, C. The Emerging Landscape of Single-Molecule Protein Sequencing Technologies. *Nat Methods* **2021**, *18* (6), 604–617. https://doi.org/10.1038/s41592-021-01143-1.

(2) Xue, L.; Yamazaki, H.; Ren, R.; Wanunu, M.; Ivanov, A. P.; Edel, J. B. Solid-State Nanopore Sensors. *Nat. Rev. Mater.* **2020**, *5* (12), 931–951. https://doi.org/10.1038/s41578-020-0229-6.

(3) Leggett, R. M.; Clark, M. D. A World of Opportunities with Nanopore Sequencing. *Journal of Experimental Botany* **2017**, *68* (20), 5419–5429. https://doi.org/10.1093/jxb/erx289.

(4) Kumar Sharma, R.; Agrawal, I.; Dai, L.; Doyle, P. S.; Garaj, S. Complex DNA Knots Detected with a Nanopore Sensor. *Nat Commun* **2019**, *10* (1), 4473. https://doi.org/10.1038/s41467-019-12358-4.

(5) Plesa, C.; Verschueren, D.; Pud, S.; Van Der Torre, J.; Ruitenberg, J. W.; Witteveen, M. J.; Jonsson, M. P.; Grosberg, A. Y.; Rabin, Y.; Dekker, C. Direct Observation of DNA Knots Using a Solid-State Nanopore. *Nature Nanotech* **2016**, *11* (12), 1093–1097. https://doi.org/10.1038/nnano.2016.153.

(6) Mihovilovic, M.; Hagerty, N.; Stein, D. Statistics of DNA Capture by a Solid-State Nanopore. *Phys. Rev. Lett.* **2013**, *110* (2), 028102. https://doi.org/10.1103/PhysRevLett.110.028102.

(7) Chau, C.; Marcuccio, F.; Soulias, D.; Edwards, M. A.; Tuplin, A.; Radford, S. E.; Hewitt, E.; Actis, P. Probing RNA Conformations Using a Polymer–Electrolyte Solid-State Nanopore. *ACS Nano* **2022**, *16* (12), 20075–20085. https://doi.org/10.1021/acsnano.2c08312.

(8) Namani, S.; Kavetsky, K.; Lin, C.-Y.; Maharjan, S.; Gamper, H. B.; Li, N.-S.; Piccirilli, J. A.; Hou, Y.-M.; Drndic, M. Unraveling RNA Conformation Dynamics in Mitochondrial Encephalomyopathy, Lactic Acidosis, and Stroke-like Episode Syndrome with Solid-State Nanopores. *ACS Nano* **2024**, *18* (26), 17240–17250. https://doi.org/10.1021/acsnano.4c04625.

(9) Tripathi, P.; Chandler, M.; Maffeo, C. M.; Fallahi, A.; Makhamreh, A.; Halman, J.; Aksimentiev, A.; Afonin, K. A.; Wanunu, M. Discrimination of RNA Fiber Structures Using Solid-State Nanopores. *Nanoscale* **2022**, *14* (18), 6866–6875. https://doi.org/10.1039/D1NR08002D.

(10) Niedzwiecki, D. J.; Grazul, J.; Movileanu, L. Single-Molecule Observation of Protein Adsorption onto an Inorganic Surface. *J. Am. Chem. Soc.* **2010**, *132* (31), 10816–10822. https://doi.org/10.1021/ja1026858.

(11) Talaga, D. S.; Li, J. Single-Molecule Protein Unfolding in Solid State Nanopores. *J. Am. Chem. Soc.* **2009**, *131* (26), 9287–9297. https://doi.org/10.1021/ja901088b.

(12) Larkin, J.; Henley, R. Y.; Muthukumar, M.; Rosenstein, J. K.; Wanunu, M. High-Bandwidth Protein Analysis Using Solid-State Nanopores. *Biophysical Journal* **2014**, *106* (3), 696–704. https://doi.org/10.1016/j.bpj.2013.12.025.

(13) Yusko, E. C.; Bruhn, B. R.; Eggenberger, O. M.; Houghtaling, J.; Rollings, R. C.; Walsh, N. C.; Nandivada, S.; Pindrus, M.; Hall, A. R.; Sept, D.; Li, J.; Kalonia, D. S.; Mayer, M. Real-Time Shape Approximation and Fingerprinting of Single Proteins Using a Nanopore. *Nature Nanotech* **2017**, *12* (4), 360–367. https://doi.org/10.1038/nnano.2016.267.

(14) Niedzwiecki, D. J.; Movileanu, L. Monitoring Protein Adsorption with Solid-State Nanopores. *JoVE* **2011**, No. 58, 3560. https://doi.org/10.3791/3560.

(15) Niedzwiecki, D. J.; Lanci, C. J.; Shemer, G.; Cheng, P. S.; Saven, J. G.; Drndić, M. Observing Changes in the Structure and Oligomerization State of a Helical Protein Dimer Using Solid-State Nanopores. *ACS Nano* **2015**, *9* (9), 8907–8915. https://doi.org/10.1021/acsnano.5b02714.

(16) Oukhaled, A.; Cressiot, B.; Bacri, L.; Pastoriza-Gallego, M.; Betton, J.-M.; Bourhis, E.; Jede, R.; Gierak, J.; Auvray, L.; Pelta, J. Dynamics of Completely Unfolded and Native Proteins through Solid-State Nanopores as a Function of Electric Driving Force. *ACS Nano* **2011**, *5* (5), 3628–3638. https://doi.org/10.1021/nn1034795.

(17) Waduge, P.; Hu, R.; Bandarkar, P.; Yamazaki, H.; Cressiot, B.; Zhao, Q.; Whitford, P. C.; Wanunu, M. Nanopore-Based Measurements of Protein Size, Fluctuations, and Conformational Changes. *ACS Nano* **2017**, *11* (6), 5706–5716. https://doi.org/10.1021/acsnano.7b01212.

(18) Shekar, S.; Niedzwiecki, D. J.; Chien, C.-C.; Ong, P.; Fleischer, D. A.; Lin, J.; Rosenstein, J. K.; Drndić, M.; Shepard, K. L. Measurement of DNA Translocation Dynamics in a Solid-State Nanopore at 100 Ns Temporal Resolution. *Nano Lett.* **2016**, *16* (7), 4483–4489. https://doi.org/10.1021/acs.nanolett.6b01661.

(19) Lin, C.-Y.; Fotis, R.; Xia, Z.; Kavetsky, K.; Chou, Y.-C.; Niedzwiecki, D. J.; Biondi, M.; Thei, F.; Drndić, M. Ultrafast Polymer Dynamics through a Nanopore. *Nano Lett.* **2022**, *22* (21), 8719–8727. https://doi.org/10.1021/acs.nanolett.2c03546.

(20) Plesa, C.; Kowalczyk, S. W.; Zinsmeester, R.; Grosberg, A. Y.; Rabin, Y.; Dekker, C. Fast Translocation of Proteins through Solid State Nanopores. *Nano Lett.* **2013**, *13* (2), 658–663. https://doi.org/10.1021/nl3042678.

(21) Hüttemann, M.; Pecina, P.; Rainbolt, M.; Sanderson, T. H.; Kagan, V. E.; Samavati, L.; Doan, J. W.; Lee, I. The Multiple Functions of Cytochrome c and Their Regulation in Life and Death Decisions of the Mammalian Cell: From Respiration to Apoptosis. *Mitochondrion* **2011**, *11* (3), 369–381. https://doi.org/10.1016/j.mito.2011.01.010.

(22) Tripathi, P.; Benabbas, A.; Mehrafrooz, B.; Yamazaki, H.; Aksimentiev, A.; Champion, P. M.; Wanunu, M. Electrical Unfolding of Cytochrome *c* during Translocation through a Nanopore Constriction. *Proc. Natl. Acad. Sci. U.S.A.* **2021**, *118* (17), e2016262118. https://doi.org/10.1073/pnas.2016262118.

(23) Tripathi, P.; Firouzbakht, A.; Gruebele, M.; Wanunu, M. Direct Observation of Single-Protein Transition State Passage by Nanopore Ionic Current Jumps. *J. Phys. Chem. Lett.* **2022**, *13* (25), 5918–5924. https://doi.org/10.1021/acs.jpclett.2c01009.

(24) Meng, E. C.; Goddard, T. D.; Pettersen, E. F.; Couch, G. S.; Pearson, Z. J.; Morris, J. H.; Ferrin, T. E. UCSF CHIMERAX: Tools for Structure Building and Analysis. *Protein Science* **2023**, *32* (11), e4792. https://doi.org/10.1002/pro.4792.

(25) Chien, C.-C.; Shekar, S.; Niedzwiecki, D. J.; Shepard, K. L.; Drndić, M. Single-Stranded DNA Translocation Recordings through Solid-State Nanopores on Glass Chips at 10 MHz Measurement Bandwidth. *ACS Nano* **2019**, *13* (9), 10545–10554. https://doi.org/10.1021/acsnano.9b04626.

(26) Colombini, M. Voltage Gating in the Mitochondrial Channel, VDAC. *J. Membrain Biol.* **1989**, *111* (2), 103–111. https://doi.org/10.1007/BF01871775.

(27) Schreiber, J.; Karplus, K. Analysis of Nanopore Data Using Hidden Markov Models. *Bioinformatics* **2015**, *31* (12), 1897–1903. https://doi.org/10.1093/bioinformatics/btv046.

(28) Liu, S.-C.; Ying, Y.-L.; Li, W.-H.; Wan, Y.-J.; Long, Y.-T. Snapshotting the Transient Conformations and Tracing the Multiple Pathways of Single Peptide Folding Using a Solid-State Nanopore. *Chem. Sci.* **2021**, *12* (9), 3282–3289. https://doi.org/10.1039/D0SC06106A.

(29) Timp, W.; Comer, J.; Aksimentiev, A. DNA Base-Calling from a Nanopore Using a Viterbi Algorithm. *Biophysical Journal* **2012**, *102* (10), L37–L39. https://doi.org/10.1016/j.bpj.2012.04.009.

(30) Kowalczyk, S. W.; Grosberg, A. Y.; Rabin, Y.; Dekker, C. Modeling the Conductance and DNA Blockade of Solid-State Nanopores. *Nanotechnology* **2011**, *22* (31), 315101. https://doi.org/10.1088/0957-4484/22/31/315101.

(31) Chou, Y.-C.; Chen, J.; Lin, C.-Y.; Drndić, M. Engineering Adjustable Two-Pore Devices for Parallel Ion Transport and DNA Translocations. *J. Chem. Phys.* **2021**, *154* (10), 105102. https://doi.org/10.1063/5.0044227.

(32) Qiu, Y.; Lin, C.-Y.; Hinkle, P.; Plett, T. S.; Yang, C.; Chacko, J. V.; Digman, M. A.; Yeh, L.-H.; Hsu, J.-P.; Siwy, Z. S. Highly Charged Particles Cause a Larger Current Blockage in Micropores Compared to Neutral Particles. *ACS Nano* **2016**, *10* (9), 8413–8422. https://doi.org/10.1021/acsnano.6b03280.